\documentclass[journal=jacsat,manuscript=article,layout=twocolumn]{achemso}

\usepackage{chemformula} 
\usepackage[T1]{fontenc} 
\usepackage{braket}
\usepackage{amsmath}
\usepackage{hyperref}
\usepackage{epstopdf}

\author{Henning Kirchberg}
\affiliation{I. Institut f\"ur Theoretische Physik, Universit\"at Hamburg, Jungiusstr.\ 9, 20355 Hamburg, Germany}
\email{henning.kirchberg@physik.uni-hamburg.de}
\author{Michael Thorwart}
\affiliation{I. Institut f\"ur Theoretische Physik, Universit\"at Hamburg, Jungiusstr.\ 9, 20355 Hamburg, Germany}
\author{Abraham Nitzan}
\affiliation{Department of Chemistry, University of Pennsylvania, Philadelphia, Pennsylvania 19104, United States of America}
\email{anitzan@sas.upenn.edu}

\title[Charge Transfer in Non-Equilibrated Solvents]
  {Charge Transfer Through Redox Molecular Junctions in Non-Equilibrated Solvents}

\begin{document}

\begin{abstract}
Molecular conduction operating in dielectric solvent environments are often described using kinetic rates based on Marcus theory of electron transfer at a molecule-metal electrode interface. However, the successive nature of charge transfer in such system implies that the solvent does not necessarily reach equilibrium in such process. Here we generalize the theory to account for solvent nonequilibrium and consider a molecular junction consisting of an electronic donor-acceptor system coupled to two metallic electrodes and placed in a polarizable solvent. We determine the nonequilbrium distribution of the solvent by solving diffusion equations in the strong- and weak-friction limits and calculate the charge current and its fluctuating behavior. In extreme limits: the absence of the solvent or fast solvent relaxation, the charge transfer statistics is Poissonian, while it becomes correlated by the dynamic solvent in between these limits. A Kramers-like turnover of the nonequilibrium current as a function of the solvent damping is found. Finally, we propose a way to tune the solvent-induced damping using geometrical control of the solvent dielectric response in nanostructured solvent channels.
\end{abstract}


\section{}
\textit{Introduction.-} Metal-molecule-metal junctions that operate in dielectric solvent are common in many applications, for example in junctions controlled by electrochemical gating. Charge transport in such junctions often occurs by successive electron hopping between molecular sites as well as between the molecule and the metal leads \cite{zha2008}. In the simplest model when the molecule supports only one electron localization site, this site is repeatedly occupied and de-occupied as electrons hop between the metal and the molecule. Each such hopping event is accompanied by solvent relaxation (so called reorganization) to accommodate the molecule's charging state and determines the time for transient localization. The overall conduction in this case is determined by metal-molecule coupling, the solvent-imparted stabilization (determined by the reorganization energy or the polaron formation energy) and solvent fluctuations needed to overcome the localization barrier. 
 
Theoretical treatment of such sequential hopping events usually rely on the Marcus electron transfer theory \cite{zha2008}. This is a transition-state type theory that assumes that solvent relaxation between hopping events proceeds to full thermal equilibrium so that the next electron transfer event takes place out of a thermal  equilibrium distribution of solvent configurations. Such behavior can be realized when solvent reorganization in response to charge localization on the connecting molecule is fast relative to the molecule-leads' tunneling rates. The other extreme limit, where the solvent is not sensitive to the molecular charge redistribution, corresponds to co-tunneling transport that is described by  the Landauer theory. However, even when localization and solvent relaxation make consecutive hopping, the dominant mechanism, the assumption of full thermal relaxation embedded in the Marcus theory, is not necessarily valid, and an extension of Marcus theory to situations where the electron transfer rate is 'solvent controlled' in the sense that it depends on the solvent relaxation dynamics, is  abundant \cite{zus1980,zus1995,mar1986,kos1986}.

Most relevant to the present work are studies that focus on solvent dynamic effects in bridge mediated electron transfer \cite{sum2001,sai2009}, where solvent dynamics is not manifested just in the electron-transfer rate, but also in the determination of the very nature of the processes between the limiting forms of co-tunneling (or super-exchange) and sequential hopping.

The papers cited above are representative of a substantial body of literature that analyzes deviations of the electron transfer rate from its transition state theory (TST) form due to finite solvent relaxation time. TST becomes valid when this relaxation is assumed fast, implying that bridge mediated transfer is sequential, with the individual hopping rates assuming their respective Marcus form. As discussed extensively in the context of the classical theory of barrier crossing, slow relaxation does not necessarily imply weak coupling to the solvent \cite{kra1940,nit2007,han1990}. Indeed, solvent relaxation in response to solute state-change is manifested in classical barrier crossing rates as a crossover between the low- and high-friction limits \cite{han1990,mel1986,pol1989}, with the rate increasing or decreasing with increasing friction at the low- (underdamped) and high- (overdamped) friction limits, while the TST approximation provides an upper bound on the rate in the intermediate crossover region. Most treatments of such effects in electron transfer have focused on the low-friction case although the other limit has also been considered \cite{oka2000}.

In this paper, we consider the effect of finite solvent-induced relaxation on hopping conduction in molecular junctions, thus going beyond the standard treatments of transport in solvated molecular junctions that rely on Marcus theory. We keep other assumptions of this theory: electron tunneling is conditional on the possibility to conserve the total (electronic and nuclear) energy, and broadening of molecular electronic levels due to their interaction with the metal leads is disregarded so that transfer rates into or out of the metal electron energy level at $\epsilon$ are proportional to $\Gamma f(\epsilon)$ and $\Gamma[1-f(\epsilon)]$, respectively. Here $\Gamma$ is a golden rule rate associated with the molecule-metal coupling, and $f(\epsilon)$ is the Fermi function \cite{mar1964,NitzanBook}. The corresponding Marcus rates $k_{AB}$ (from the electrode to the molecule) and the reverse $k_{BA}$ are
\begin{align}
\label{eqMarcus1}
k_{AB} &= \int d \epsilon \Gamma(\epsilon) f(\epsilon) F(E_{AB}+\epsilon) \\ 
k_{BA} &= \int d \epsilon \Gamma(\epsilon)[1-f(\epsilon)] F(E_{BA}-\epsilon),
\label{eqMarcus2}
\end{align}
where $E_{AB}=E_A-E_B$ is the energy difference between the molecular states $A$ and $B$ and 
\begin{align}
\label{eqFraCon}
F(x)=\frac{1}{\sqrt{4\pi E_R k_B T}}\exp{-\frac{(x-E_R)^2}{4E_R k_B T}},
\end{align}

with $T$ and $k_B$ denoting the temperature and the Boltzmann constant, respectively (throughout this work we assume that the temperature of the metal electrodes and the solvent are equal) and $E_R$ is the solvent reorganization energy - the free energy released by relaxation of the solvent to its stable (equilibrium) configuration following a sudden transition between the oxidized and reduced molecular states. 

Eq.\ (\ref{eqFraCon}) is the high-temperature limit of the average, over a thermal (Boltzmann) distribution of solvent configurations in the initial electronic state, of the Franck-Condon factors that reflect the difference between the solvent equilibrium configurations in the two molecular electronic states. When the finite timescale of the solvent relaxation is taken into account, this thermal distribution is replaced by a time-dependent distribution that reflects this relaxation. Below we describe this dynamics by diffusion (Smoluchowski) equations that take different forms in the high- and low-friction limit. In analogy to the Kramers theory of activated barrier crossing \cite{kra1940}, the high-friction (overdamped) limit is characterized by diffusion along the position of the reaction coordinate, while the low-friction limit is described by diffusion in energy space. In either case, the average junction current resulting from this calculation depends on the friction parameter $\gamma$ that determines the solvent relaxation rate. Of further significance is the dependence of the current noise on this relaxation rate. For small damping, the process is Markovian and the hopping statistics is Poissonian. When solvent relaxation occurs on a finite timescale, successive electron hopping events are correlated. We suggest that a combined measurement of the average charge current and the current noise may serve to identify such situations in solvated electronic junctions.
Finally, we discuss the possibility to realize these limiting behaviors as well as intermediate solvent friction behaviors through the dependence of solvent relaxation dynamics on its geometrical confinement.

\section{}
\textit{Theoretical Model.-} We consider a molecule weakly coupled to two metal electrodes $R$ (right) and $L$ (left) which are modeled as free electron reservoirs characterized by their respective chemical $\mu_K$, and electrical, $\phi_{K}$ ($K=L,R$), potentials due the externally applied voltage.

The associated Fermi functions for the electron energy $\epsilon$ read
\begin{align}
f_K(\epsilon)=\frac{1}{\exp\left(\frac{\epsilon-\mu_K+e\Phi_K}{k_BT}\right)+1},
\label{eq1}
\end{align}
where $K=L,R$, and $e$, $k_B$ and $T$ are the electron charge, the Boltzmann constant and the temperature, respectively.

The molecule comprises a single transport channel, thus forming a two-state system with an oxidized state $A$ with $N-1$ electrons, and energy $E_A$, and a reduced state $B$ with $N$ electrons, and energy $E_B$. 
The molecule is furthermore embedded in polar solvent that imposes a fluctuating environment and responds electrodynamically to the charging state of the molecule \cite{lev}. 

In the Marcus theory  \cite{mar1956a,mar1956b} this response is expressed by a distribution of solvent configurations along a single reaction coordinate, $x$, determined by free energy surfaces that depend on the molecular electronic state according to \cite{NitzanBook}.
\begin{align}
E_A(x,\epsilon)&=E_A+\frac{1}{2} \hbar  \omega_0 x^2+\epsilon \label{eq2}\\ 
E_B(x)&=E_B+\frac{1}{2} \hbar \omega_0(x-d)^2.
\label{eq3}
\end{align}
In this shifted harmonic surfaces model, $E_A$ and $E_B$ are the electronic energies at the equilibrium solvent configurations, chosen as $x_A=0$ and $x_B=d$ for the state $A$ and $B$ respectively. 
The harmonic forms and the identical curvatures of these surfaces correspond to the assumption that the solvent responds linearly to the charging state of the molecule and has the consequence that the reorganization energy
\begin{align}\label{reorgen}
E_R=\frac{1}{2}\hbar \omega_0 d^2
\end{align}
is the same irrespective on the process direction from $A$ to $B$ or vice versa. 

The single electron energy $\epsilon$ is added to the molecular energy in state $A$ (cf. Eq.\ (\ref{eq1})), expressing the fact that when the molecule is oxidized, the electron removed from it is occupying a single electron state of energy $\epsilon$ in the metal. The integrals over $\epsilon$ in Eqs.\ (\ref{eqMarcus1}) and (\ref{eqMarcus2}) reflect the broad band of single electron states in the metal. The Marcus theory makes two further simplifying assumptions: the small molecule-metal coupling and the high-temperature limit. Together they have several implications. First, the assumption that $\hbar \Gamma \ll k_BT$, allows to disregard level broadening due to the finite lifetime of excess electron or holes on the molecule. Secondly, in the high-temperature limit nuclear tunneling can be disregarded in evaluating the electron hopping rate, namely such events are assumed to be dominated by crossings of nuclear potential surfaces. Finally, the small electronic coupling $\Gamma$ makes it possible to use perturbation theory to the lowest order in the electronic coupling for evaluating the electron hopping probability, essentially disregarding level splitting in this calculation (e.g., using the non-adiabatic limit of the Landau-Zenner expression for this probability). Under these assumptions, electron transfer events are dominated by nuclear configurations where $E_A(x)=E_B(x)$, namely at the transition point along the reaction coordinate given by
\begin{align}
x_{TR}=\frac{E_B-E_A-\epsilon + \frac{1}{2} \hbar \omega_0 d^2}{\hbar \omega_0 d}\, .
\label{eq4}
\end{align}

Marcus theory \cite{mar1956a,mar1956b} is based on transition state theory, an essentially equilibrium theory of reaction rates. It provides a framework for representing the solvent state on a one dimensional free energy surface defined with respect to a single reaction coordinate which is valid provided that the electronic energies in states $A$ and $B$ depend only on a single solvent quantifier, in the present case the local solvent polarization. Extending this theory to the dynamical regime requires additional assumption, tacitly made in most studies like those cited above, that the same free energy surfaces $E_A(x)$ and $E_B(x)$ can be used as potential energy surfaces for the reaction coordinate $x$, provided that account is taken for the fact that this coordinate can exchange energy with all other solvent degrees of freedom. Making the additional assumption that this dynamics is Markovian, the motion of the reaction coordinate when the molecule is in state $A$ or $B$ can be described using a Langevin equation obtained by augmenting Newtonian dynamics on potential surfaces (\ref{eq2}) or (\ref{eq3}) by a Stokes friction of strength $\gamma$ and a corresponding random noise that together satisfy the standard fluctuation-dissipation relation. Equivalently, the probability distribution for the position and velocity of the reaction coordinate $P_j(x,v;t)$ for the molecular state $j=A,B$, obeys under these conditions the Fokker-Planck equation
\begin{align}
\label{eqFokker}
\frac{\partial P_j (x,v;t)}{\partial t}=&\omega_0 \frac{d \bar{V}_j}{d x}\frac{\partial P_j}{\partial v}-\omega_0 v \frac{\partial P_j}{\partial x} \\ \notag &+\gamma \bigg [ \frac{\partial}{\partial v}(vP_j)+\frac{k_B T}{\hbar \omega_0} \frac{\partial^2 P_j}{\partial v^2} \bigg ]\, .
\end{align}
In Eq.\ (\ref{eqFokker}), the normalized potential surfaces are $\bar{V}_j=V_j/(\hbar \omega_0)$ with $V_A(x)=\frac{1}{2}\hbar \omega_0 x^2$ and $V_B(x)=\frac{1}{2}\hbar \omega_0 (x-d)^2$. Note that the position and velocity variables in Eqs.\ (\ref{eq2}),(\ref{eq3}) and (\ref{eqFokker}) are dimensionless. The solvent properties that enter at this level of description are manifested via the parameters $\omega_0$ and $\gamma$ that can be obtained from fitting of the observed dielectric response of the solvent to standard dielectric response models \cite{MayBook}. The solvent-molecule coupling enters via the parameter $d$ that determines the solvent reorganization energy $E_R$ as given in Eq.\ (\ref{reorgen}). Note that the overdamped limit of Eq.\ (\ref{eqFokker}) has been used in the Zusman generalization of the Marcus theory \cite{zus1980,zus1995}.

In the following, in analogy to Kramers' treatment of activated barrier crossing \cite{kra1940}, we consider the implications of this dynamics in two limits. In the overdamped limit, $\gamma \gg \omega_0$, Eq.\ (\ref{eqFokker}) leads to a Smoluchowski equation, that describes diffusion along the $x$ coordinate, 
\begin{align}
\label{eqFokker1}
\frac{\partial P_j (x,t)}{\partial t} =\frac{\omega_0}{\hbar \beta \gamma} \frac{\partial}{\partial x} \bigg [ \frac{\partial}{\partial x}+\beta \hbar \omega_0 \frac{d \bar{V}_j}{d x} \bigg] P_j(x,t)\, ,
\end{align}
where $\beta=(k_BT)^{-1}$. In the opposite underdamped limit, $\gamma \ll \omega_0$, the relaxation implied by Eq.\ (\ref{eqFokker}) may be reduced, after phase averaging, to diffusion in energy space, which is described by 
\begin{align}
\label{eqFokker2}
\frac{\partial P_j (E,t)}{\partial t} =\frac{\partial}{\partial E} \bigg [ \gamma E \bigg[1+k_B T \frac{\partial}{\partial E}\bigg] P_j(E,t)\bigg]\, .
\end{align}

The distribution functions $P(x,t)$ in Eq.\ (\ref{eqFokker1}) or $P(E,t)$ in Eq.\ (\ref{eqFokker2}) replace the Boltzmann distribution in evaluating the instantaneous probability for electron transfer in the Marcus theory, leading to time-dependent rates that will replace the rates given by Eqs.\ (\ref{eqMarcus1}) and (\ref{eqMarcus2}). We note that the stationary solution of both Eqs.\ (\ref{eqFokker1}) and (\ref{eqFokker2}) is the Boltzmann distribution, implying that transition state theory will be recovered when relaxation is fast, $\gamma \to 0$ in Eq.\ (\ref{eqFokker1}), or $\gamma\to \infty$ in Eq.\ (\ref{eqFokker2}). 

In what follows, using Eqs.\ (\ref{eqFokker1}) and (\ref{eqFokker2}) as our starting points, we construct numerical simulation procedures for calculating the charge-transport characteristics operating in solvent environments in the corresponding dynamical limits (see Sec.\ S3 in Supporting Information for details). We investigate their implications for standard observables like the average charge current and the charge current noise, as function of voltage bias and solvent induced friction. 
\section{}
\textit{Overdamped regime.-}
\begin{figure}[!h]
  \centering
  \includegraphics[width=0.45\textwidth]{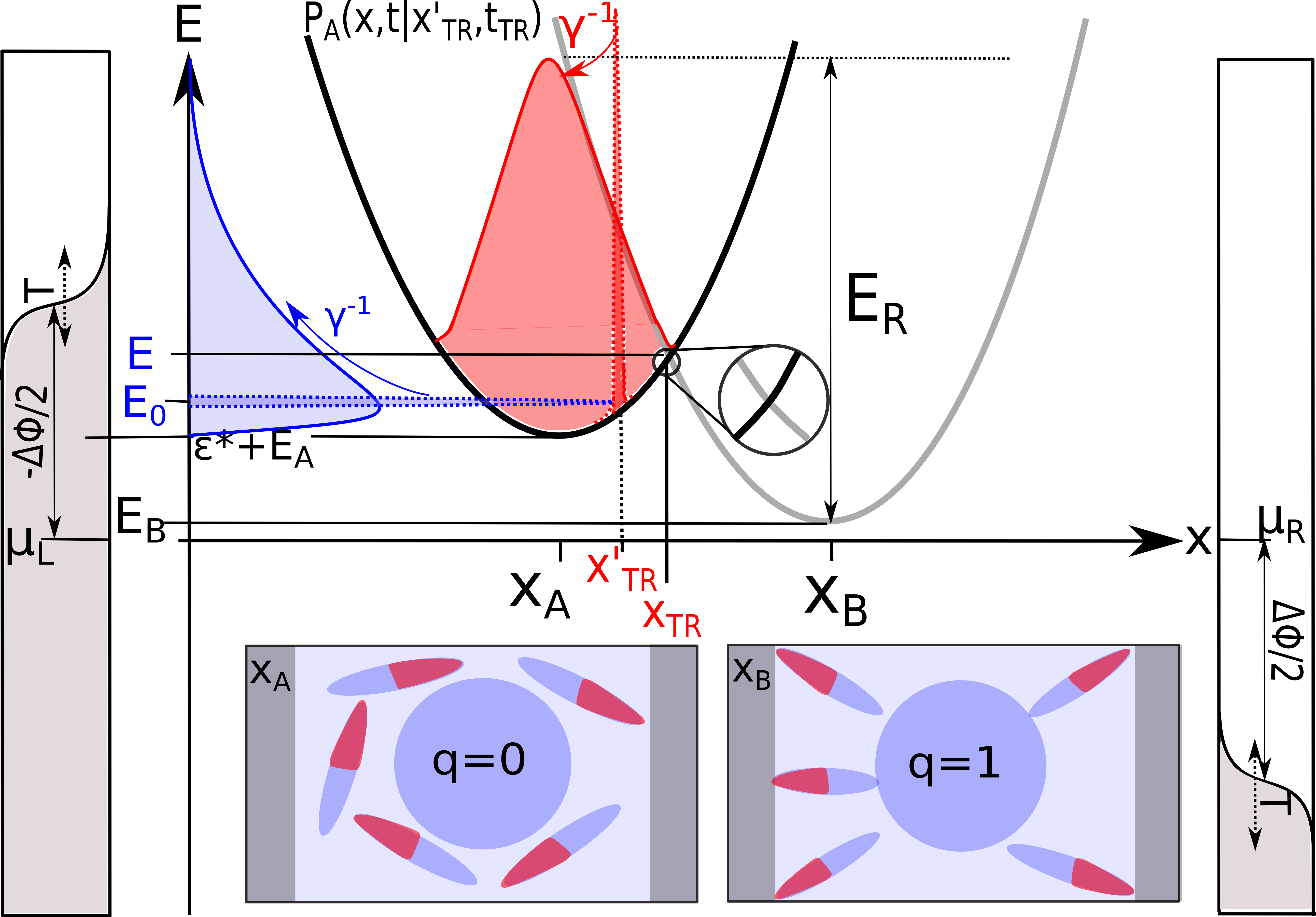}
  \caption{\label{fig1}Energy representation of the nonequilibrium electron transfer process. The black diabat represents the oxidized state $A$ including the energy alignment with a charge of arbitrarily chosen energy $\epsilon^*$, the grey diabat corresponds to the reduced state $B$. The energy distribution of the charges and vacancies in the left ($K=L$) and right ($K=R$) electrode follows from the Fermi distribution $f_K(\epsilon)$ (or $1-f_K(\epsilon)$, respectively, Eq.\ (\ref{eq1})) with temperature $T$. The state energy strongly depends on the solvent configuration described by the reaction coordinate $x$ (see Eqs.\ (\ref{eq2}) and (\ref{eq3})). The relaxation of the reaction coordinate $x$ after a previous transition at $x'_{TR}$ is described by its probability density (red colored) under the action of the damping $\gamma$. In the strong friction limit the velocity quickly settles to its equilibrium distribution and the dynamics is affected by the distribution of the position coordinate (red colored) while in the weak friction case, averaging over phase leaves energy, whose distribution is displayed in blue, as the relevant dynamical variable. The boxes show the equilibrium solvent configuration $x_A=0$ with equilibrium energy $E_A$ for the oxidized state and $x_B=d$ for the reduced state with equilibrium energy $E_B$ indicating the fluctuating dipoles of the solvent molecules.}
\end{figure}
The probability densities resulting from the Smoluchowski equations (\ref{eqFokker1}) for the overdamped reaction coordinate $x$ ($\gamma \gg \omega_0$) can be calculated exactly (see Sec.\ S1 in Supporting Information for details). We are interested in the time evolution of the probability density $P(x,t|x'_{TR},t_{TR})$ that the reaction coordinate takes the value $x$ following a previous transition event that took place at time $t_{TR}$ at position $x'_{TR}$ of this coordinate. This corresponds to the initial condition $P(x,t_{TR}|x'_{TR},t_{TR})=\delta(x-x'_{TR})$ for which we find the evolutions in the state $A$ and $B$ (see also illustration in Fig.\ \ref{fig1}) 
\begin{align}
\label{eq8} 
&P_A(x,t|x'_{TR},t_{TR})=\sqrt{\frac{D}{2\pi [1-a^2(t-t_{TR})]}} \\ \notag &\times \exp\left\{-\frac{D}{2}\frac{[x-a(t-t_{TR})x'_{TR}]^2}{1-a^2(t-t_{TR})}\right\} \, ,
\end{align}
\begin{align}
\label{eq9}
&P_B(x,t|x'_{TR},t_{TR})=\sqrt{\frac{D}{2\pi  [1-a^2(t-t_{TR})]}} \\ \notag &\times \exp\left\{-\frac{D}{2}\frac{[x-d-a(t-t_{TR})(x'_{TR}-d)]^2}{1-a^2(t-t_{TR})}\right\},
\end{align}
where $D = \beta \hbar \omega_0$ and $a(t)=\exp\left(-\frac{\omega_0^2}{\gamma}t\right)$. For $\gamma\to0$, $a(t)\to \infty$ for all time $t>0$, indication "instantaneous" relaxation to an equilibrium Boltzmann distribution in the corresponding wells.

$P_{A}(x,t|x'_{TR},t_{TR})dx$ is the probability to find a solvent configuration with a reaction coordinate in $[x,x+dx]$ for the the oxidized state $A$ at time $t$, given that the previous transition from the reduced state $B$ has occurred at the solvent configuration $x'_{TR}$ at time $t_{TR}$. Correspondingly, $P_{B}(x,t|x'_{TR},t_{TR})dx$ describes the equivalent for the reduced state $B$. It is important to notice that the next electronic transition can take place at any $x$. This $x$ then becomes the next transition configuration $x_{TR}$ for which the energy $\epsilon(x_{TR})$ and consequently the probabilities to find a corresponding metal level occupied $f_K(x)$ or unoccupied $1-f_K(x)$ are determined from Eq.\ (\ref{eq4}). 

Correspondingly, the ET rates (probabilities per unit time), $k_{AB}$ into the molecule, and $k_{BA}$ out of the molecule, are given in this high-friction limit by
\begin{align}
\label{eq6}
k_{AB}^{K}&(t-t_{TR};x'_{TR}) \\ \notag &= \Gamma \int_{-\infty}^{\infty} dx P_A(x,t|x'_{TR},t_{TR})f_K(x) \\ \noindent 
\label{eq7}
k_{BA}^{K}&(t-t_{TR};x'_{TR})  \\ \notag &= \Gamma \int_{-\infty}^{\infty} dx P_B(x,t|x'_{TR},t_{TR})\left[1-f_K(x)\right]\, ,
\end{align}
where $K=L$ or $R$.  $\Gamma$ is assumed to be independent of the solvent configuration $x$, while $\Gamma^{-1}$ characterizes the time span between the electronic hopping events. 

We emphasize again that the integration over all solvent configuration $x$ in  Eqs.\ (\ref{eq6}) and (\ref{eq7}) can be extended to $\pm \infty$ because a transition may occur at every solvent configuration along the reaction coordinate, subjected to the Pauli principle that is accounted for explicitly in Eqs.\ (\ref{eq6}) and (\ref{eq7}). In agreement with the above observation that the limit $\gamma \to 0$ correspond to "infinitely fast" relaxation to equilibrium, the rates given in Eqs.\ (\ref{eq6}) and (\ref{eq7}) become the thermal Marcus rates of Eqs.\  (\ref{eqMarcus1}) and (\ref{eqMarcus2}).

Using the rates of Eqs.\ (\ref{eq6}) and (\ref{eq7}), we can calculate the average charge current with a numerical Monte Carlo procedure (see Sec. S3 in Supporting Information for details). We set $\mu_R=\mu_L=\Delta E=E_B-E_A$ in Eqs.\ (\ref{eq1}) to (\ref{eq3}) and apply a symmetric bias voltage $\Phi_R=-\Phi_L=\Delta \Phi/2$  between the leads. The solvent dynamics is propagated using Eq.\ (\ref{eq8}) or (\ref{eq9}) depending on the present state of the molecule, and at any time step transition is attempted using the probabilities (\ref{eq6}) or (\ref{eq7}) depending again on the current molecular state.

\begin{figure}[h!]
\centering
\includegraphics[width=\linewidth]{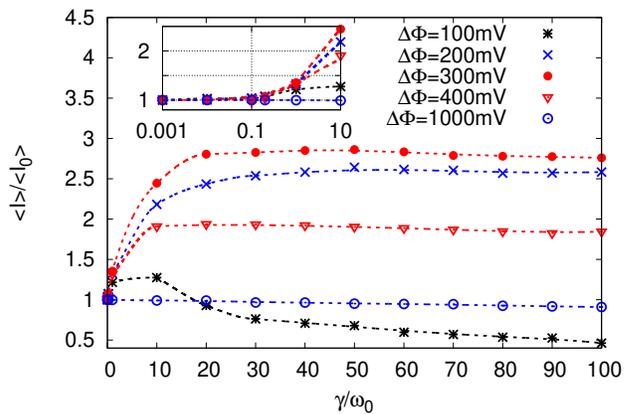}
\caption{Expectation value of the current $\langle I \rangle/\langle I_0 \rangle$ at ambient temperature $T=300\rm K$ for the reorganization energy $E_R=200 \rm meV$ at different applied bias potentials $\Delta \Phi$ as indicated as a function of the damping rate $\gamma$. The current is normalized to $\langle I_0 \rangle$ for infinitely fast solvent relaxation, $\gamma \to 0$. The dotted lines are functional fits to the data points for better readability. The inset shows a zoom to the regime of small $\gamma$.}
\label{fig9a}
\end{figure}
The average charge current  $\langle I \rangle$ obtained from this calculation is shown in Fig.\ \ref{fig9a} as a function of the bias voltage and the solvent induced damping $\gamma$. Depending on the voltage bias, three modes of behavior are seen. (a) When the applied bias voltage is moderately larger than the thermal energy, i.e. $e\Delta \Phi/2>k_BT$ (while $k_BT\sim 25 \rm meV$ at $300 \rm K$) the current increases, then saturates, with increasing $\gamma$ (see Fig. \ref{fig9a} for 200-400 $\rm mV$). The reason for this behavior is that for large $\gamma$ relaxation is slow, therefore, the system remains close to a transition configuration, i.e., energy barrier to the transition does not develop quickly enough. 
(b) For considerably larger bias (1000 $\rm mV$ in Fig.\ \ref{fig9a}) there is no sensitivity to solvent configuration (states are vacant on one lead and occupied on the other for any configuration), hence no effect on the average current at which this configuration evolves is seen.
(c) For small bias (100 $\rm mV$ in Fig.\ \ref{fig9a}) increasing $\gamma$ initially enhances the electronic current for the same reason as in (a): At larger $\gamma$ staying in the small energy window at which transfer can occur implies a larger probability of an electron transfer. However, some relaxation is needed to stabilize the "product" of this transfer. Further increase of $\gamma$ makes such relaxation too slow and leads to current reduction because for such large $\gamma$ the dot level sticks in an energy region which large probability to return to the electrode where it originated. 

Reiterating, a larger friction $\gamma$ implies, in the limit under consideration, a slower solvent-induced stabilization of the electron on the molecular bridge. 
The initial increase of the current with growing $\gamma$ for relatively small damping for all imposed biases (see inset Fig.\ \ref{fig9a}) may be understood as solvent-mediated delay near the molecular transition state configuration. For larger $\gamma$, this delay becomes a practical freezing of configuration that is more ($e\Delta \Phi/2>k_BT$) or less ($e\Delta \Phi/2\sim k_B T$) favorable to subsequent electron transfer events.

We note that, although the diffusion equation in the overdamped regime is strictly valid only for $\gamma \gg \omega_0$, it reproduces the Boltzmann distribution as a solution for $\gamma=0$. Yet, the calculated average current in the intermediate regime $\gamma \sim \omega_0$ may be seen as an interpolation between the tractable limits of vanishing and strong damping. 

\begin{figure}[h!]
\centering
\includegraphics[width=\linewidth]{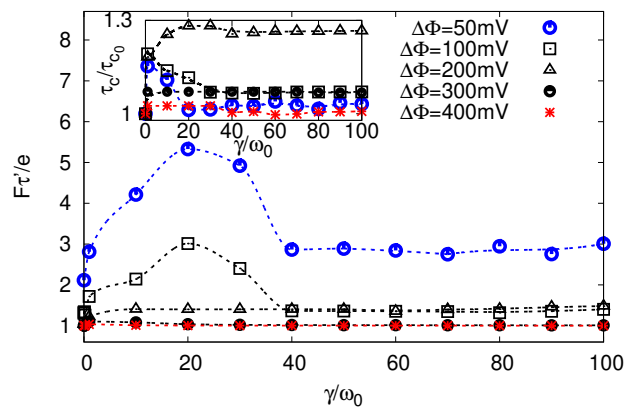}
\caption{Normalized Fano factor $F \tau'/e$ at ambient temperature $T=300\rm K$ for different applied bias potentials $\Delta \Phi$ at reorganization energy $E_R=200 \rm meV$ by varying the damping rate $\gamma$. $\tau'$ is the observation time (see Eq.\ (S52) in the Supporting Information) and $e$ the electron charge. Inset: Normalized correlation time $\tau_c/\tau_{c_0}$ by varying the damping $\gamma$ from fitting $C(t)=\langle I(t)I(0)\rangle-\langle I\rangle^2$ (see Eq.\ (S50) in the Supporting Information) to $\exp[-t/\tau_c]$. $\tau_{c_0}$ is the correlation time for infinitely fast solvent relaxation, $\gamma \to 0$}. The dotted lines are functional fits to the data points for better readability.
\label{fig9c}
\end{figure}

We next consider the current noise (see Eq.\ (S52) of Sec. S3 in Supporting Information). As commonly done, we characterize the noise in terms of the Fano factor \cite{ren2007,fan1947}, $F=\frac{\langle I^2\rangle-\langle I\rangle^2}{\langle I\rangle}>1$, with the limit $F=1$ achieved for Poissonian (uncorrelated) statistics. The result, displayed in Fig.\ \ref{fig9c}, shows
clear evidence of a non-Poissonian ET process for a non-vanishing friction, while Poisson statistics characterizes the limit $\gamma \to 0$ (fast solvent-induced relaxation). Once the applied bias voltage is large enough, the solvent relaxation shows no impact on the ET because the position of the dot ''level'' does not change its orientation relative to empty and occupied single electron states of the metal, hence $F=1$ for all $\gamma$. 
For an applied bias in the regime of relevant thermal fluctuations, i.e., $e\Delta\Phi \sim k_BT$ (see $\Delta \Phi =50 \rm mV$ and $\Delta \Phi =100 \rm mV$ in Fig.\ \ref{fig9c}), the Fano factor first increases before it declines to a steady value when the molecular ''state'' is localized in a region when its sees probabilities of similar magnitudes for vacancies or occupations on the leads of both sides. 

More insight on current correlations may be obtained from the correlation function $C(t)=\langle I(t)I(0) \rangle - \langle I\rangle^2$ (see Eq.\ (S50) of Sec. S3 in Supporting Information). The inset in Fig.\ \ref{fig9c} shows the correlation time $\tau_c$, defined by fitting $C(t)$ to an exponential $C(t)\approx\exp[-\frac{t}{\tau_c}]$. An increased correlation time $\tau_c$ with larger damping signals a higher current correlation in correspondence with the observations made above. 

\section{}
\textit{The low-damping regime.-} In low-damping regime, $\gamma \ll \omega_0$, the reaction coordinate oscillates many times under its harmonic restoring force before appreciable relaxation occurs. In this case, as in the Kramers \cite{kra1940} regime of low damping, the parameter that determines the probability of electron transfer and relaxes due to the interaction with the thermal environment is the energy. In the Kramers case, the rate in this limit is determined by the inverse mean first-passage time to reach a critical energy. Here, the rate can be obtained from the golden rule, say for the $A \to B$ transition, in the form
\begin{align}
\label{eq10}
k_{AB}= \frac{2\pi}{\hbar} & \int_{-\infty}^{\infty} d \epsilon \rho_M(\epsilon) f(\epsilon)\sum_{v;v'}|\braket{v|\hat{V}_{A,B}|v'}|^2 \\ \notag & \times \delta(E_A-E_B+\epsilon+E_{b}(v)-E_{b}(v'))\, ,
\end{align}
where $E_A(v)$, $E_B(v')$ are eigenvalues of the solvent Hamiltonian (harmonic oscillator state on the potential surfaces $E_A(x)$ and $E_B(x)$), $\hat{V}_{AB}$ is the interstate coupling and $\rho_m(\epsilon)$ is the density of single electron states in the metal. We further consider the density of states $\rho_M(\epsilon)=\rho_M=$ const.\ as independent of the electron energy.

In Sec. S4 of the Supporting Information, we show that Eq.\ (\ref{eq10}) can be transformed into the expression
\begin{align}
\label{eq10b}
k_{AB}=& \Gamma \int_{-\infty}^{\infty} d \epsilon f(\epsilon) \int_{-\infty}^{\infty} dt e^{i(E_A-E_B+\epsilon)t/\hbar} \\ \notag &\times e^{-ig^2\sin(\omega_0t)+g^2[2n(t)+1][\cos(\omega_0t)-1]},
\end{align}
where $\Gamma=\frac{\rho_M|V_{A,B}|^2}{\hbar^2}$ is the conditional rate and $g$ determines the coupling between the solvent (oscillator) states and the molecular state $A$ or $B$. 
Eqs.\ (\ref{eq10}) and (\ref{eq10b}) are not limited to the semi-classical limit used in the Marcus theory and, in fact constitute, for this low-damping limit, an extension of the Marcus formalism. Keeping, for simplicity, the Marcus level of treatment which entails an additional approximation equivalent to the high-temperature limit used in the transition state theory limit of this treatment (see Ref.\  \cite{NitzanBook}, Chapter 16): Assuming  $n(t) \gg 1$ in the relevant range of solvent energy, and because the reaction coordinate $x$ interacts in this energy range with many solvent degrees of freedom, the integrand is very short-lived \cite{jor1976} and can be approximated by expanding the sine term up to first and the cosine term up to second order in the argument $\omega_0 t$. This short-time expansion leads to
\begin{align}
\label{eq35}
&k_{AB}= \Gamma \int_{-\infty}^{\infty} d \epsilon  f(\epsilon) \\ \notag & \times \int_{-\infty}^{\infty} dt e^{i(E_A-E_B+\epsilon)t/\hbar -it g^2 \omega_0-t^2  g^2 \omega_0^2[2n(t)+1]/2} \\ \notag
=&\Gamma \sqrt{\frac{\pi}{a}} \int_{-\infty}^{\infty} d \epsilon  f(\epsilon) e^{-\frac{(E_A-E_B+\epsilon-E_R)^2}{4\hbar^2 a}},
\end{align}
where we consider $n(t)$ to evolve only slightly in time for slow energy relaxation ($\gamma\ll \omega_0$). Moreover, we perform the time integral in Eq.\ (\ref{eq35}) by regarding $n(t)$ as a constant parameter. We introduce $a=(2n(t)+1)g^2\omega_0^2/2 \simeq E(t)E_R/ {\hbar}^2$ as well as the reorganization energy $E_R=\hbar g^2\omega_0$ and the solvent energy $E(t)=\hbar \omega_0 n(t)$ in Eq.\ (\ref{eq35}). 

The rates for the electron insertion and removal process accompanied by slow energy relaxation finally follow as 
\begin{align}
\label{eq36}
k_{AB}(t;E|E_0)
&=\Gamma \sqrt{\frac{\pi}{E(t)E_{R}}} \\ \notag & \int_{-\infty}^{\infty} d \epsilon f(\epsilon) e^{-\frac{(E_A-E_B+\epsilon-E_R)^2}{4E(t)E_R}}\\
\label{eq37}
k_{BA}(t;E|E_0)&
=\Gamma \sqrt{\frac{\pi}{E(t)E_{R}}} \\ \notag & \int_{-\infty}^{\infty} d \epsilon[1-f(\epsilon)] e^{-\frac{(E_B-E_A-\epsilon-E_R)^2}{4E(t)E_R}}.
\end{align}
Remarkably, the final result under this short-time approximation is similar to Marcus' result, except that the thermal energy $k_BT$ is replaced by $E(t)$ - the (time-dependent) solvent energy expressed by the energy content in the reaction coordinate, calculated at time $t$, which is the time elapsed since the proceeding electron has hopped onto or out off the molecule, and subject to the initial condition $E(t=0)=E_0$, which is the energy at which the proceeding hopping took place. 
\begin{figure}[h!]
\centering
\includegraphics[width=\linewidth]{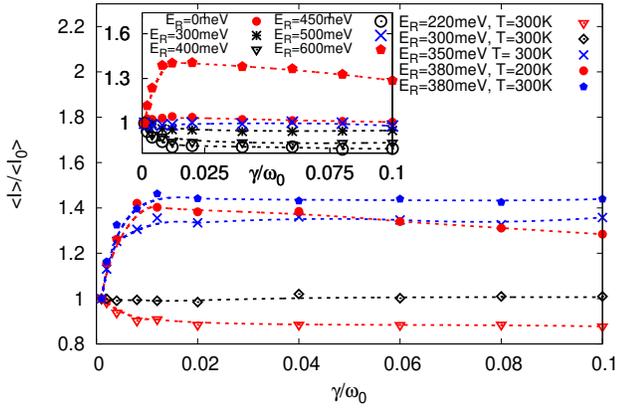}
\caption{Expectation value of the current $\langle I\rangle/\langle I_0\rangle$ for different values of the reorganization energy $E_R$ for varying damping strength $\gamma$ in the underdamped regime. The bias potential is set to $\Delta \Phi=600\rm mV$. The current is normalized to $\langle I_0\rangle$ obtained for $\gamma \to 0$ (no solvent energy relaxation). Inset: $\langle I\rangle/\langle I_0\rangle$ for different values of the reorganization energy $E_R$ at the bias voltage $\Delta \Phi=1000\rm mV$ at temperature $T=300\rm K$. The dotted lines are functional fits to the data points for better readability.}
\label{fig10}
\end{figure}

Next, we consider the energy relaxation (or diffusion) dynamics, sketched in Fig.\ \ref{fig1}.
As before, we look at the time evolution between two electron hopping events. For definiteness, assume that the next electron hopping will be an $A\to B$ transition and the proceeding $B\to A$ event happened at time $t_0$ under system energy $E_0$. Following this event, the probability to find the system at time $t$ with energy $E$ is the solution of Eq.\ (\ref{eqFokker2}) with the initial condition $P(E,t_0|E_0,t_0)=\delta(E-E_0)$. It is given by (see Sec. S2 in Supporting Information for details)
\begin{align}
\label{eq10c}
P(&E,t|E_0,t_0)=\frac{1}{k_B T [1-e^{-\gamma (t-t_0)}]} \\\notag & \times \exp\bigg[\frac{-[E_0 e^{-\gamma (t-t_0)} +E]}{k_B T [1-e^{-\gamma (t-t_0)}]}\bigg] \\ \notag & \times\sum_{m=0}^\infty \frac{\big[\frac{EE_0e^{-\gamma (t-t_0)}}{k^2_B T^2[1-e^{-\gamma (t-t_0)}]^2}\big ]^m}{m!^2}\Theta(E)\, .
\end{align}

Using the energy dependent rates of Eqs.\ (\ref{eq36}) and (\ref{eq37}) and the evolving probability distribution for this energy, Eq.\ (\ref{eq10c}), we have constructed a numerical Monte Carlo procedure for generating a sequence of electronic transitions (see Sec. S3 in Supporting Information for details) from which the average current as well as the current noise can be evaluated. For the results shown in Fig.\ \ref{fig10} we have used $\mu_R=\mu_L=\Delta E=E_B-E_A$, see Eqs.\ (\ref{eq1}) to (\ref{eq3}), and a symmetric bias voltage $\Phi_R=-\Phi_L=\Delta \Phi/2$ in Eq.\ (\ref{eq1}).

Consider first the average current and recall that in the low-damping regime energy relaxation is faster for larger $\gamma$. For $e\Delta\Phi/2<E_R$, the average current increases with growing solvent-induced damping $\gamma$ (see Fig.\ \ref{fig10} for $E_R=380 \rm meV$ and $E_R=350 \rm meV$). This increase appears to stem from the fact that the energy distribution becomes broader in time, such that more metal states can be accessed. This enhances the rates and hence the current through the molecule for growing $\gamma$. It is evident that this effect becomes more pronounced at higher temperature (see Fig.\ \ref{fig10} for $E_R=380 \rm meV$ at $T=200\rm K$ and $T=300 \rm K$) since the thermal fluctuations are enhanced, thus leading to thermally activated ET processes. If the reorganization energy becomes much larger than the applied bias voltage, the broadening of the molecular states becomes irrelevant since the alignment with the Fermi states of the leads vanishes.
When the bias voltage $\Delta \Phi$ satisfies $e\Delta\Phi/2=E_R$ (see Fig.\ \ref{fig10} for $E_R=300 \rm meV$ or inset of Fig.\ \ref{fig10} for $E_R=500 \rm meV$), the average current shows no sensitivity to an increasing damping $\gamma$. The reason appears to be the fact that in this case the molecular energy state aligns with the occupied/vacant electronic energy levels in the leads.

For $e\Delta\Phi/2>E_R$, the average current decreases with increasing solvent induced damping $\gamma$ (see Fig.\ \ref{fig10} for $E_R=220 \rm meV$). In this case, the molecular states which are broadened due to $\gamma$ drop out of the conduction window of the leads. This reduces the ET rates (Eqs.\ (\ref{eq36}) and (\ref{eq37})) and, thus, the current through the molecule. 

\begin{figure}[h!]
\centering
\includegraphics[width=\linewidth]{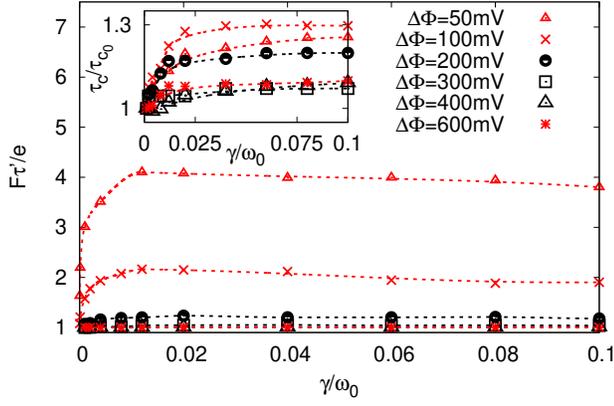}
\caption{Normalized Fano factor $F\tau'/e$ at ambient temperature $T=300\rm K$ for the reorganization energy $E_R=200 \rm meV$ at different applied bias voltages $\Delta \Phi$ as a function of $\gamma$. Here, $\tau'$ is the observation time (see Eq.\ (S52) of Sec. S3 in Supporting Information) and $e$ the electron charge. Inset: Normalized correlation time $\tau_c/\tau_{c_0}$ for varying damping $\gamma$ as obtained from the fitting $C(t)=\langle I(t)I(0)\rangle-\langle I\rangle^2$ (see Eq.\ (S50) of Sec. S3 in Supporting Information) to $\exp[-t/\tau_c]$. Here, $\tau_{c_0}$ is the correlation time for $\gamma \to 0$ (no solvent energy relaxation)}. The dotted lines are functional fits to the data points for better readability.
\label{fig11}
\end{figure}

In addition to the average current, we also consider the noise properties (see Eq.\ (S52) of Sec. S3 in Supporting Information). Figure \ref{fig11} shows the calculated Fano factor $F=\frac{\langle I^2\rangle-\langle I\rangle^2}{\langle I\rangle}>1$.  It reveals a strongly non-Poissonian ET process for a small applied bias voltage calculated for a large reorganization energy. Here, the effect of molecular level broadening induced by larger $\gamma$ leads to correlated electron hopping in time just as the current through the molecule. This correlation disappears if the applied voltage exceeds the reorganization energy considerably. 
Again by fitting an exponential $C(t)\propto \exp[-\frac{t}{\tau_c}]$ to the normalized calculated correlation function the impact of damping on the current correlations can be directly quantified (see inset in Fig.\ \ref{fig11}). An increased correlation time $\tau_c$ with an enhanced damping directly leads to a stronger current correlation and confirms the prior observations.  

\section{}
\textit{Kramers-like turnover.-}
\begin{figure}[h!]
\centering
\includegraphics[width=\linewidth]{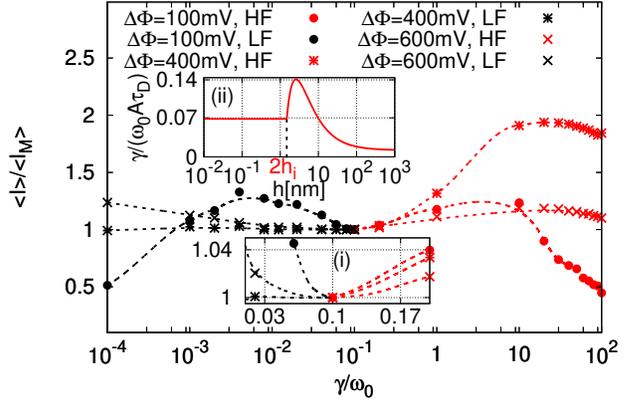}
\caption{The ratio $\langle I\rangle/\langle I_M\rangle$ between the actual average current $\langle I \rangle$ and the current obtained in the TST (Marcus) limit ($\gamma \to 0$ in the high-friction calculation (red dots) or $\gamma \to \infty$ in the low-friction one (black dots)), plotted as a function of $\gamma$ in a range encompassing both low- and high-friction (LF and HF, respectively) regimes and displayed for different values of the bias potential $\Delta \Phi$. The reorganization energy is set to $E_R=200\rm meV$ and the temperature is taken to $T=300\rm K$. In the main panel, the results displayed in red were obtained from the high-friction formalism based on Eqs.\ (\ref{eq6}) and (\ref{eq7}) while those in black correspond to the low-friction calculation of Eqs.\ (\ref{eq36}) and (\ref{eq37}). The dotted lines are functional fits to the data points for better readability. Inset i: Focus on the transition from low to high friction of the average current. Inset ii: Effective frequency-dependent damping $\gamma /\omega_0$ in confined nanochannels of height $h$ filled with water ($\hbar\equiv 1$) based on data taken from Fumagalli \textit{et al.} \cite{fum2018}.}
\label{fig12}
\end{figure}
It is interesting to consider the dependence of the average current on the damping strength. This is shown in Fig.\ \ref{fig12}. For $e\Delta \Phi/2 < E_R$, we see for  $\Delta\Phi=100\rm mV$ in the low-friction regime that $\langle I\rangle\propto \gamma$ and that the current is controlled by energy diffusion and the resulting broadening of the molecular level inside the conduction window. In the high-damping regime,  we find that $\langle I\rangle \propto \gamma^{-1}$ for small  bias voltage, $e\Delta \Phi/2 \sim k_BT$, and the current is strongly influenced by the configurational change along the reaction coordinate. For large damping, the solvent configurations freeze the molecular state at a transition state which may fall into occupied or vacant electronic levels in the leads of opposite directions. This reduces the current. For $e\Delta \Phi/2 > E_R$, see Fig.\ \ref{fig12} for $\Delta\Phi=600\rm mV$, the molecular level broadening exceeds the width of the conduction window for small but growing $\gamma$ such that $\langle I\rangle \propto \gamma^{-1}$ while for $e\Delta \Phi/2 > k_B T$ the solvent stabilizes a possible transition state in the high-damping regime where $\langle I\rangle \propto \gamma$.

The 'turnover' between the regimes of low and the high damping seen in Fig.\ \ref{fig12} is reminiscent of the Kramers turnover of activated barrier crossing rates \cite{kra1940}. Despite a conceptual similarity, there is an important difference:  We consider the average current which is composed in the sequential hopping affected by four different time-dependent ET rates describing electron insertion  and removal to/from the molecular dot via the left/right contact. All four rates depend on the solvent damping and the applied bias voltage. Therefore, there is no  one-to-one mapping of the charge current to the Kramers rate as shown in Chapter 14 of Ref.\ \cite{NitzanBook} in the strict sense. Yet, the analogy is obvious. 
\section{}
\textit{Proposal for experimental control of solvent damping.-}
The question may arise how to suitably tune the damping strength $\gamma$ in a real junction experiment. Since the solvent adjusts to the charge distribution on the molecule, its relaxation properties are determined by electromagnetic response. In particular, for a polar solvent, they depend via the spectral density linearly on the dielectric function \cite{MayBook}. Provided that the damping experienced by the Marcus reaction coordinate stems from the solvent, we may aim to control damping by tuning the dielectric properties of the solvent. This is indeed  possible in a dielectric solvent confined on the nanoscale and can be realized by tuning the geometry of the confinement. A proof-of-principle is the recent observation of the dependence of the dielectric function of water confined in nanochannels \cite{fum2018}. 

To illustrate this connection in more detail, we use the spectral density of a fluctuating dipolar solvent specified by \cite{MayBook}
\begin{align}
J(\omega)=\frac{2 E_R}{\pi \hbar} \frac{\omega_D}{\omega}\frac{1}{\omega^2+\omega_D^2},
\end{align}
where $\omega_D=\epsilon_0/(\epsilon_\infty \tau_D)$ with the low- ($\epsilon_0$) and high- ($\epsilon_\infty$) frequency dielectric constants and the Debye relaxation time $\tau_D$ for solvent relaxation described by an Onsager continuum model of the solvent with Debye relaxation. In the Ohmic regime where $\omega_D \gg \omega_0$, the damping kernel for the Langevin equation for the diffusive coordinates $x$ or $E$ can be evaluated as $\tilde{\gamma}(t)=\Theta(t)\omega_0\int_0^\infty d\omega J(\omega) \omega \cos(\omega t)=\Theta(t)\omega_0\frac{E_R}{\hbar}e^{-\omega_D t}=\omega_0\frac{\epsilon_\infty}{\epsilon_0} \frac{E_R \tau_D}{\hbar}\delta(t)=\gamma\delta(t)$. For the latter equation, we use the definition of the damping kernel of Ref.\ \cite{WeissBook} , but multiply $J(\omega)$ by $\omega^2$ (we note the different definition of the spectral densities of  Refs.\ \cite{MayBook} and \cite{WeissBook}) and the limit $\omega_D\to\infty$ to obtain the  $\delta$-distribution.

It is interesting to see that Fumagalli \textit{et al.} \cite{fum2018} have found experimentally an anomalous decline of the static dielectric constant following the phenomenological relation $\epsilon_0(h)=h/[2h_i/\epsilon_i+(h-2h_i)/\epsilon_{bulk}]$ for water confined in nanochannels of height $h$ with $\epsilon_i=2.1$, $\epsilon_\infty=1.8$ and $h_i=7.4$ {\AA}. They explain the strong reduction of $\epsilon_0(h)$ for the nanostructured water channel as compared to the bulk configuration with a restricted mobility of the water dipoles at the boundary surfaces which the nanochannel forms with the host material in which they are immersed. 

The relationship between the damping constant and the reorganization energy implies in turn that the friction experienced by a solute in a dielectric solvent depends on the dielectric function. For example, using the common relation $E_R=A\big[\frac{1}{\epsilon_\infty}-\frac{1}{\epsilon_0}\big]$ \cite{MayBook}, $A$ being a solvent/solute dependent coefficient with the dimension of an energy, we find a connection between $\gamma$ and $\epsilon_0$ and consequently between $\gamma$ and any geometrical parameter that may affect $\epsilon_0$. In detail, we find a geometry-dependent damping strength 
\begin{align}
\gamma(h)=\left[\frac{1}{\epsilon_\infty}-\frac{1}{\epsilon_0(h)}\right]\omega_0\frac{\epsilon_\infty}{\epsilon_0(h)} \frac{A \tau_D}{\hbar}\, ,
\end{align}
where the strong change of the static dielectric constant in confined geometries enters. 

Obviously, we can effectively tune the damping strength $\gamma$ by tuning the height $h$ of the nanochannel, i.e., the degree of confinement of the solvent, in which the metal-molecule-metal junction operates. The resulting dependence of $\gamma$ on  $h$ is shown in the inset (ii) of Fig.\ \ref{fig12}. This effect intensifies further if one takes into account the enhanced relaxation time $\tau_D$ for water molecules restricted in their mobility \cite{tie2010}. For different modes of operation, i.e., for different relations of $E_R$, $\Delta\Phi$ and $k_BT$, we thus may control the current $\langle I\rangle$ by tuning $\gamma$. Even though the change of $\gamma$ may not lead to a turnover from low to high damping or vice versa, the characteristic $\langle I\rangle\propto \gamma$ or $\langle I\rangle\propto\gamma^{-1}$ indicates the high- or low-damping regime. 

\section{}
\textit{Conclusion.-}We have calculated the average charge current for a sequential electron transfer in a metal-molecule-metal junction, where consecutive events of charge transfer my be strongly influenced by relaxation of surrounding solvent, thereby affecting the observed current and its fluctuations. The resulting dynamics was described by a kinetic model that comprises rates of electron exchange between molecules and leads with relaxation of the thermal environment in response to the changing molecular charge. This relaxation is described by a diffusion process that focuses on the reaction coordinate in the high-friction regime and on its associated energy in the low-friction limit.  
We have considered the average current as well as the current noise. In the regime of low friction, an increasing solvent damping strength leads to an increasingly fluctuating energy level which is associated to an effective broadening of the molecular energy level that lies inside ($e\Delta \Phi/2 < E_R$) or outside ($e\Delta \Phi/2 > E_R$) the conduction window of the metal-molecular-metal junction  (using the language of a combined Marcus-Landauer model \cite{mig2012,yua2018}). Consequently, the average current is enhanced ($e\Delta \Phi/2 < E_R$) or reduced ($e\Delta \Phi/2 > E_R$) when friction becomes larger in this regime. 
In the regime of large damping, increasing friction implies a more slowly relaxing configuration along the reaction coordinate. Therefore, in the course of successive electron transfer events the molecular configuration remains more localized near a transition point. This leads to an average current that increases with friction when $e\Delta \Phi/2 > k_BT$, but decreases when an enhanced localization leads to an increased probability of back ET, which becomes possible when $e\Delta \Phi/2 \sim k_BT$. A unified look at both regimes indicates a Kramers-like turnover of electron transfer which translates into a corresponding turnover behavior of the overall current as a function of the damping strength. This clearly reflects the nonequilibrium fluctuations at work under an applied bias voltage.

It should be noted that "turnover" is a generic mode of behavior of rate (or transport) processes. The overall transport dynamics determined by the underlying rate(s) changes its character between different regimes by varying some control parameter. 
In the Kramer problem this parameter is the friction on the reaction coordinate while in our case it is the (not unrelated) dielectric relaxation associated with the solvation coordinate (as defined by Marcus). Other mechanisms have been considered, see, e.g. Refs.\ \cite{cao2013,dubi2018} , where the control parameter changes the dephasing of local energy levels leading to turnover in exciton transport behavior. To the best of our knowledge, our present work is the first where the consequence of such turnover is considered not only for the average rate (or transport coefficient) but also for the noise in the ensuing current.

The calculated Fano factor indicates a non-Poissonian current statistics for an enhanced damping in both regimes for a small applied bias voltage and  a large reorganization energy. Once the applied voltage is large enough, the solvent-induced friction has no impact on the ET, because the solvent-mediated orientation (in the high-damping regime) or the broadening (in the low-damping regime) of the molecular level does not change relative to empty and occupied single-electron states in the metal. Current noise as an additional observable may help to distinguish between different modes of operation. Additionally, this mechanism may help to interpret current-voltage observations (see Ref.\ \cite{fun2019}). 

Furthermore, we illustrate a viable means to tune the damping strength $\gamma$ of the solvent which depends on the dielectric properties of the solvent via the static dielectric constant. For this, we illustrate a possible way of how to effectively use  experimentally measured data which report an extreme decline of  the static dielectric constant of water in confined nanochannels of variable height. By this, we deduce a direct dependence of $\gamma$ on the height of the nanochannel which can be readily used to observe and control the characteristic current-damping behavior in molecular junctions. 

\begin{acknowledgement}
\\
We acknowledge financial support by the Deutsche Forschungsgemeinschaft within a German-Israeli Research Project (project number 320285192). A.N. also acknowledges support  from the U.S. National Science Foundation (Grant No. CHE1665291).

\end{acknowledgement}

\begin{suppinfo}
In the Supporting Information, we first derive in Sec.\ S1 the exact solution of the Smoluchowski equation for the probability distribution of the reaction coordinate $x$ in the regime of high damping and for the energy $E$ for low damping in Sec.\ S2. With the resulting probability distributions, we are able to determine ET rates which explicitly depend on the time elapsed between successive electron transfer processes. We use this rate to set up a Monte Carlo procedure for the sequential insertion and removal of one electron to calculate the average current under nonequilibrium conditions in Sec.\ S3. Finally, in Sec.\ S4, we show the formal derivation of Eq.\ (\ref{eq10b}). \\
\end{suppinfo}

\end{document}